# Trends in smart lighting for the Internet of Things

Jorge Higuera[1], Aleix Llenas[1,2], Josep Carreras [2],

Catalonia Institute for Energy Research (IREC)[1], Barcelona, Spain

Ledmotive Technologies[2], Barcelona, Spain

**Abstract**

Smart lighting is an underlying concept that links three main aspects: solid state lighting (SSL) technologies, advanced control and universal communication interfaces following global standards. However, this conceptualization is constantly evolving to comply with the guidelines of the next generation of devices that work in the Internet of Things (IoT) ecosystem.

Modern smart lighting systems are based on light emitting diode (LED) technology and involve advanced drivers that have features such as dynamic spectral light reproduction and advanced sensing capabilities. The ultimate feature is of additional advanced services serving as hub for optical communications that allows coexistence with traditional Wi-Fi gateways in indoor environments. In this context, lighting systems are evolving to support different wireless communications interfaces compatible with the IoT ecosystem.

Market tendencies of SSL systems predict the accelerated expansion of connected IoT lighting control systems in different markets from smart homes and industrial environments. These systems offer advanced features never seen before such as advanced spectral control of the light source and also, the inclusion of several communication interfaces. These are mainly wired, radiofrequencies (RF) and optical wireless communications (OWC) interfaces for advanced services such as sensing and visible light communications (VLC). In this chapter we present how to design and realize IoT-based smart lighting systems for different applications using different IoT-centric lighting architectures. Finally, different standards and aspects related to interoperability and web services are explained taking into account commercial smart lighting platforms.

1. **Introduction**

Smart lighting has received a lot of global attention recently as it is an evolution from human centric lighting with light emitting diodes (LEDs) [1],[2] and/or organic light emitting diodes (OLEDs) [3] for dual use, illumination and communication purposes. This technology aims to fuse drivers with sensing, control algorithms and wireless communications to deploy scalable lighting solutions able to work autonomously in the Internet of Things (IoT) ecosystem.

A typical explanation for smart lighting is a lighting system containing energy efficient LED drivers, miniaturized digital lighting sensors, advanced control algorithms and standardized communication interfaces to cooperate and interact in a lighting network.

At its core, a smart lighting system is being conceived as an adaptable lighting system with the objective to improve visual comfort, as well energy efficiency. Diverse implementations of smart lighting systems involve different communication interfaces and additional capabilities such as light spectral reproduction in real-time, advanced detection options with illuminance sensors, color sensors and micro-spectrometers.

The most advanced lighting systems include additional features beyond conventional illumination purposes. These capabilities are related to spectrally-tunable features, optical communications and extended control capabilities that open a new paradigm associated with the possibility to remotely control the system from the Internet and mimic any light spectrum imaginable in real-time.

Solid state lighting (SSL) fixtures are generally efficient solutions for home and industrial purposes. In these systems, the luminous efficacy of LEDs is increasing every year, while their price is steadily decreasing. Also, according to Haitz's law predictions [4] the luminous efficacy will rise every decade and in the future the cost per lumen will continue to decrease. This prediction is being largely fulfilled due to advances in material science that predict that the amount of light generated per LED increases every year. In addition, as smart lighting systems based on LEDs involve non-polluting materials several studies show a positive life cycle analysis (LCA) [5] that makes SSL technology very attractive in current and future smart lighting systems.

From a technical outlook smart lighting solutions are far from stagnant thanks to the interest of the lighting industry in developing advanced features for the new generation of luminaires. The luminaires are specifically related to autonomous sensing capabilities, extended IoT connectivity, scalability and interoperability features in different marketable segments, and also considering optical communications as communication interface for smart lighting systems.

Potential applications for connected intelligent lighting include human-centered lighting systems, optical communications, standardized and interoperable solutions that work on

IoT networks, advanced lighting systems where specific spectral components are critical or play an important role, and can be controlled in real-time.

## 2. Smart lighting

Smart lighting is the umbrella that encompasses different solid state technologies such as LEDs and OLEDs to illuminate indoor and outdoor environments. Smart lighting systems mainly involve digital sensors, actuators drivers and communications interfaces. These lighting systems are programmed using advanced control algorithms and can be organized into lighting networks to operate remotely. Some of the most popular solutions are designed to change the light spectrum or color. They can also control the level of illumination in a room when an external event occurs, for example, when a user has been detected by an occupancy sensor or when an event occurs such as the detection of vehicles or people on a road.

The smart lighting system eliminates the need to operate the overall system in manual mode. The lighting network is programmed with an initial setup; however, each independent light can be reprogrammed to respond to the desires of people and situations throughout the day. Diverse digital communication interfaces intended for smart lighting are the digital addressable lighting interface (DALI), Ethernet, Wi-Fi, ZigBee light link or Bluetooth for the programming of predefined areas and spaces. In these systems, generally the areas are segmented depending on the people or events that may occur. This allows the systems to calculate the level of light needed, so that it can accurately calculate the levels of illuminance suitable for different tasks of the users with the advantage to calculate the power consumption in real-time.

Smart lighting systems organized as lighting networks often allow different types of lights to interact with each other, so that they can be synchronized. It is also possible to individually control a light fixture through the network by means of a remote controller, for example, with an application from a graphical interface of a mobile phone or a web browser.

**2.1 Lighting sources: white light based on LEDs**

In order to understand how solid state lighting works we need to first explain in detail how white light can be generated for illumination and/or communication purposes.

Humans are adapted to working in healthy environments that mimic the sun daylight spectrum. For this reason, we always seek to illuminate a space with white light that mimics the solar spectrum. There are two different ways to provide artificial white light with LEDs and both techniques are of interest in human centric applications (HCA and optical communications where perception of color and light spectrum are very important parameters.

The first method to obtain white light for illumination purposes, or optical communications, employs a combination of red, green and blue (RGB) LEDs [6] as shown in Figure 1.

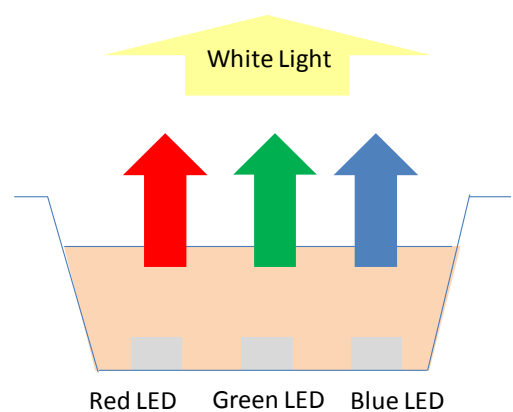

Figure 1. Creating white light by mixing individual red, green and blue LEDs

RGB LED emitters consist in three different LED color types and materials: red (aluminum gallium arsenide AlGaAs), green (indium gallium nitride InGaN) and blue (zinc selenide ZnSe) that emit jointly to obtain white light on average. The mixture of the different intensities in the LEDs are related to the obtained chromaticity of the white light, for instance, white light with more content of blue light is named as cold white light, while neutral white light involves a homogeneous mixture and warm white light indicates more relative intensity of the red emitter.

The second technique to obtain white light employs a phosphorous coating in contact with LED encapsulation resin [7] as shown in Figure 2. This method consists of inorganic blue LEDs made of indium gallium nitride (InGaN) with a phosphorous coating made of yttrium aluminum garnet (YAG) synthetic crystalline materials. This coating absorbs blue photons exciting electrons to upper orbitals from which they decay later in several steps emitting less energetic photons ranged over a wide spectrum giving rise to a either cold, or neutral, or warm white light averaged emission.

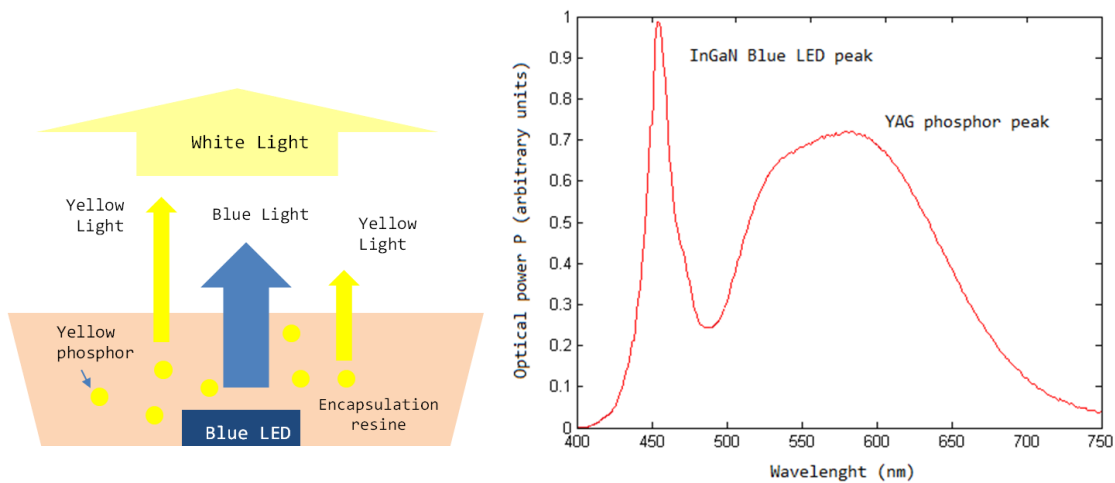

Figure 2. White light based on phosphor-converted LED. The phosphor powder is aggregated directly on the blue LED or near the LED to convert the blue light to white light

A remarkable characteristic is the fact that the mentioned coating can be either on the surface of every single blue LED or as a remote coating in an external structure, in which case there will be a cluster of blue LEDs without coating over their surfaces but with a broad general phosphorous coating at a certain distance making the same effect over all of them simultaneously.

In both cases, it is feasible to obtain different color correlated temperatures (CCT) in kelvin (K) between (2700 K to 6500 K) by changing the composition of phosphor coating or tuning different RGB LEDs to mimic the solar spectrum and also tuning the color rendering index (CRI) of the lighting source between 80 – 100. The CRI is a useful quantity to define the perceived color quality by eyes when an object is illuminated in an indoor or outdoor environment. The CRI can be changed by tuning YAG phosphorus formulation during the fabrication phase of luminaires.

## 2.2. Energy efficient LED drivers

Electronic drivers for illumination purposes are devices which regulate the power for LEDs and provide variable output current for matching light source characteristics. Most lighting systems prepared for the IoT ecosystem [8] include a power conversion stage with a constant-current LED driver, several LEDs organized in arrays and the inclusion of sensors and a communication interface as shown in Figure 3.

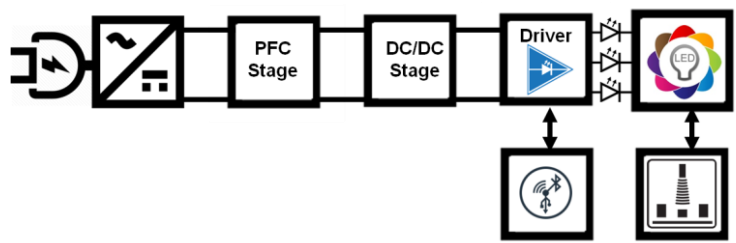

Figure 3. LED driver architecture for IoT ecosystems

The first two stages of the LED driver are related to the power supply that involve the power factor correction and the DC-DC conversion module while the third stage concerns the current control based on linear constant-current LED drivers or switching constant-current LED drivers. Boost, Buck, Buck-Boost regulators [9] or SEPIC are popular architectures for switching constant-current LED drivers.

Different types of LED drivers are capable of achieving high conversion efficiencies between 85% - 95% across a wide range of voltages. LED drivers are prone to an excess of noise generation caused by periodic switching at high frequencies when they employ pulse-width modulation (PWM) [10]. Also, to avoid visible and non-visible flicker at low frequencies, LED drivers include dynamic dimming capabilities to avoid annoying flicker while increasing the reliability of the overall system.

The stage of LEDs contains emitters are classified according to input current requirements as low power (5mA - 20mA), mid-power (30mA - 150mA), and high-power LEDs (more than 150 mA). Low power LED emitters are used in portable and low power drivers while mid-power LEDs are more useful in indoor lighting or optical communications purposes, especially in visible light communication (VLC) drivers and Li-Fi transceivers. Also, high-power LEDs are intended for LED drivers working in outdoor environments, for example in street lighting and automotive lighting. Strings of LEDs are organized in either series connection, parallel connection or a mixture of both when several LEDs are connected in series as chains, and then, several chains are connected in parallel.

## 2.3 Architectural elements

Smart lighting solutions incorporate different type of devices, systems and network types. Devices are mainly luminaires containing sensors, actuators, and advanced algorithms. A couple of advanced algorithms allow for the monitoring of daylight levels, light spectrum, or user occupation to decide a final action. The algorithms run inside devices, or to alleviate the workload of the device the algorithm can run directly in the cloud stored as a web service to send command messages to execute the different control actions.

Several algorithms for smart lighting are related to advanced operations such as to tune the color reproduction in real-time. This feature is being imposing in the market due to advances in smart lighting systems that can influence human circadian rhythms to improve mood and concentration of users. Circadian lighting systems are adopting dynamic spectral reproduction to tune the white color perception according to human rhythms in real-time. Figure 4 shows key elements for smart lighting systems.

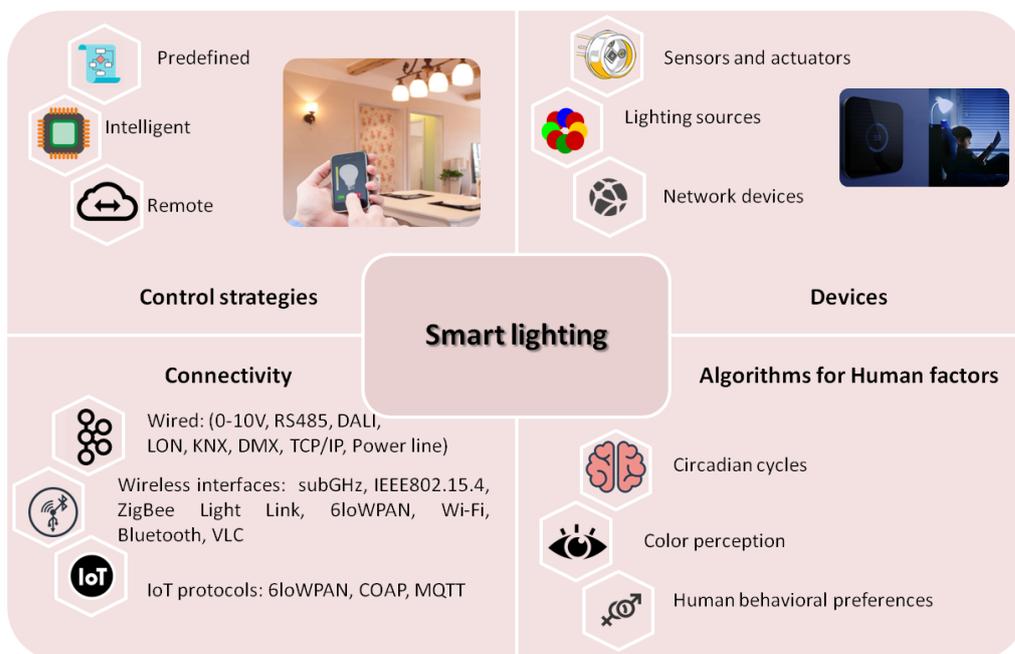

Figure 4. Key elements for smart lighting systems

Autonomous algorithms for human factors mainly try to adapt artificial lighting to human behavioral changes according to age, user preferences and gender. In the past years, it has been proven how light can modulate many non-visual functions, such as the biological clock that regulates our 24-hour circadian rhythms, our attention, hormonal secretion, body temperature and sleep. This explains why the light spectrum content, and not the color or the CCT, is the most important feature when circadian luminaires are being evaluated. Indeed, by having a control over the spectral properties through the system, it is possible to modulate and regulate many physiological properties. Besides human centric lighting, other fields like horticulture, museum lighting or performing arts are benefiting from having a complete control over the spectrum.

At system level devices may be depicted in physical and logical hierarchies to interact cooperatively in a lighting network. This network can be designed to support different physical topologies such as a ring, star, line tree, bus, mesh or a hybrid arrangement to provide a high degree of reliability. As conventional communication networks different nodes can be connected by cable or wirelessly covering different zones in a physical installation. The network management can be local or remote and would include different wired or wireless interfaces.

Different lighting products in the market are now being designed with different connectivity options to cover several wired or wireless communication interfaces to work properly in the IoT ecosystem for lighting. Common wired communication interfaces [11] use 0-10V, RS485, DALI, DMX, LON, KNX, BACnet, power line and Ethernet protocols. Besides, from the point of view of wireless standards, digital lighting luminaires incorporate some wireless communication interfaces, such as sub-GHz, IEEE802.15.4, Bluetooth, Wi-Fi, and VLC. Lighting solutions oriented to short range connectivity and low power consumption [12] are based on the physical layer IEEE 802.15.4 such as ZigBee Light Link or 6loWPAN [13].

**3. Sensors for smart lighting platforms**

At the core of a smart lighting system are different working sensor technologies and communication interfaces. The modern IoT lighting platforms aim to control lighting depending on changes in the environment with a wide range of digital sensors. Figure 5 shows different sensor types to implement in such systems.

Smart lighting involves illuminance sensors and photocells detecting small changes in illuminance levels to alert the daylight conditions. RGB color sensors are intended to detect red-green-blue content of light and tune white light in LED luminaires. For optical communications including VLC connectivity several technologies of photodiodes [14] can be used in wireless links mainly in indoor environments. Some IoT lighting platforms offer the use of passive infrared sensors (PIR) [15] for infrared radiation detection of users in movement adjusting the luminous flux according to the occupancy profile.

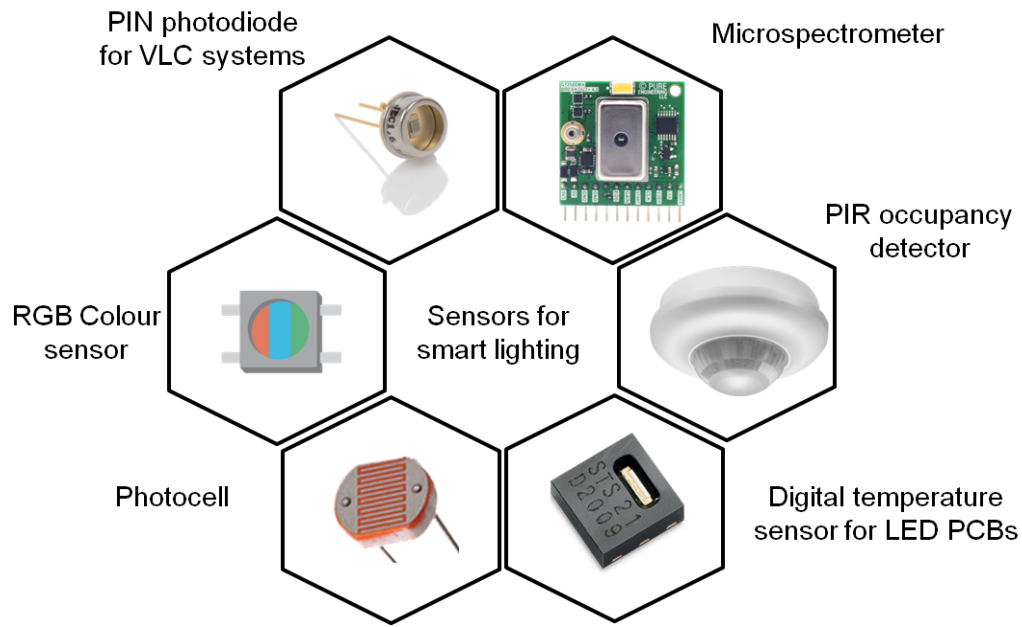

Figure 5. Sensor technologies embedded in smart lighting

In addition, more advanced functionalities such as spectral detection of light are covered with micro-spectrometers to detect the light spectrum in the visible range that can be detected by our eyes. This functionality is important in circadian lighting systems to reproduce any light spectrum and to tune CRI and CCT in real-time. In addition, it is well known that LEDs decrease their maximum luminous flux over time due mainly to temperature or aging effects, and therefore, such sensors and advanced control systems are of interest to ensure optimal performance of the lighting system.

## 3.1 Smart lighting platforms

### 3.1.1 Philips Hue platform

Philips Hue is an IoT lighting platform based on four main components: the LED light fixtures, the bridge with Internet connectivity, a web service, and apps for smart phones with an open API to send and receive commands. The Hue smart bulbs contain three types of RGB LED specially chosen to produce dynamically a range of colors and intensities. The owner of the lights can create a mesh network which enables each light to pass on messages to the next extending the range and making the system more robust. This system is connected to the Internet through a gateway that supports the stack Ethernet and the protocol ZigBee Light Link. This system uses an application program interface (API) as a tool for controlling the lights. This is a responsive RESTful interface over the HTTP protocol. The purpose of this web service interface is to give every light in the system a URL in a local network. Also, the different controllable parameters of these lights are stored as local URLs. This means that controlling the lights is achieved by simply sending a new value to this local URL.

### 3.1.2 Osram LIGHTIFY platform

Osram LIGHTIFY is an IoT platform for LED lights with Internet connectivity for homes and industrial applications. This system has scalable features and additional services, such as different interface options, supports different client profiles and has a control option via the software API based on a REST web service. The system can be controlled conveniently by a tablet computer or smartphone running the mobile app in Android or iOS operative system. The Internet connection requires a Wi-Fi router and the LIGHTIFY gateway. The gateway communicates based on the ZigBee Light Link standard, and can therefore be combined with additional ZigBee Light Link devices.

### 3.1.3 LIFX platform

LIFX is a smart light platform that uses Wi-Fi connectivity to control the system. This solution is oriented for home applications. The system includes the light bulb, the gateway and a simple-to-use app to control a single light or an entire lighting network in home.

The system has Wi-Fi connectivity and can be controlled from an Apple watch or an Android wear device. These lights work fine in several control platforms for homes such as Nest, SmartThings, Harmony, Scout Alarm etc. The system has additional features including a music interface, themes, scenes, and several control schedules.

The LIFX platform is intended to be used in indoor environments however it can be used in outdoor places that are not exposed directly to water. It can also be used in bathrooms and other areas with high moisture levels. The LIFX has a night vision feature to be used in outdoor places to work with cameras intended for security systems.

### 3.1.4 Belkin Wemo

WeMo Smart is a platform based on LED light bulbs and bridge network devices for home applications. These LED luminaires are similar to traditional incandescent bulbs however the bulbs are more energy efficient than old ones and also come with an extended life span of near to 25,000 hours.

The LED bulbs can be controlled from anywhere using a Wi-Fi network and mobile Internet. Features of this system include dimmable lights to remotely turn them on or off. Custom schedules can be defined to control each light individually or as a group in different areas of a house. The system can be turned on automatically at sunset or changed to be dimmable to watch a movie, or turn off after the user leaves the home.

The WeMo kit includes a WeMo Link device. This bridge is intended to connect up to 50 WeMo Smart LED bulbs and control them individually or as a group. The user interface is the WeMo app that is available for Android, iOS, and Kindle operating systems and works on multiple smart devices simultaneously. The WeMo app can be connected with the IFTTT web service to create extended connections with simple statements, for example to receive an SMS text or an email when the front door opens or when lights are turn on when the sun sets.

### 3.1.5 Cree connected LED bulbs

The smart lighting platform from Cree includes high quality white light fixtures that look like a traditional bulb but with an extended lifetime compared to the lifetime of about 1,000 hours for old incandescent bulbs. Cree LED bulbs can be controlled remotely and these devices are compatible with a certified ZigBee Wink hub. Every light can be synchronized with the app using any iOS or Android device. This solution is compatible with multiple home automation systems, such as Wink, SmartThings, WeMo, and ZigBee certified hub.

### 3.1.6 Flux smart LED bulbs

Flux is a Bluetooth enabled lighting system based on multicolored emitters embedded into an energy efficient LED light bulb that can be controlled with a smartphone or tablet. The light can reproduce in a color palette over 16 million colors and also different tones of white light from warm yellow white light to vibrant blue white light.

The bulbs interact with the Flux app to create personalized lighting scenes. The Flux app can change the atmosphere of a room by choosing one of the 20 pre-programmed color modes. The possibilities are defined by the user and the Flux app can control directly up to 50 smart bulbs in broadcasting mode or control each one individually.

### 3.1.7 Digital lumens

The SiteWorx platform from Digital Lumens is a wireless network system for smart lighting with embedded occupancy sensing, power metering, daylight harvesting, and progressive dimming to achieve up to 90% energy savings. This system is intended for industrial applications mainly to supervise in real-time the installation and to automatically identify power consumption and act on energy savings opportunities and to maintain safe, comfortable light levels across the facility.

The SiteWorx system has both desktop and mobile apps for iOS and Android to control single-room facilities and multi-sites based on analytics and advanced controls. The SiteWorx dashboard displays real-time energy usage and savings data, and custom reports can be created for any time period or zone.

### 3.1.8 Samsung smartThings

Samsung smart lighting modules (SLM) transform static LED luminaires in true smart digital nodes. This IoT platform incorporates advanced processing capabilities, connectivity and open architectures for collecting data and creating new applications. The core element is the SLM module that serves as the essential connector for all types of devices used in smart luminaires including sensors and advanced drivers.

Features of this system are over the air programming (OTA) and power metering in real-time. SLM modules enables devices talk to one another by supporting several wired and wireless communication protocols, utilizing not only ZigBee or DALI, but also the Bluetooth protocol. This solution can be connected to different sensor networks and effectively interacts with smart control devices such as smartphones.

### 3.1.9 Gooee ecosystem

The Gooee ecosystem is intended for sensing, control and communication with lighting devices in the cloud. This system integrates sensors to capture environmental data and human activity, along with the capability to monitor LED driver performance. The platform includes motion detection, direction, ambient light levels and temperature.

Additional sensors monitor light output, color temperature, quality and operating temperatures. Each Gooee luminaire has an on-board processor to analyze the data it receives, and to reduce the amount of data sent back to the gateway. Different algorithms determine whether collected data needs to be transmitted at the network. Gooee uses a wireless interface module to monitor power consumption, provide beacon functionality for consumer engagement and send data back to the cloud via the gateway. All data from sensors are encrypted locally and communication between WIMs and gateways is encrypted using AES128 within the mesh network. Each Gooee gateway is capable of managing up to 250 nodes but can be extended up to 1000 nodes. Every device on the Gooee network is assigned a unique ID to control and read sensing data. Gateway interaction can be conducted via the Internet, Android or an iOS app and even on the gateway itself through the embedded display. The cloud platform manages metadata to develop applications and data visualizations through its API & SDK.

### 3.1.10 Ledmotive IoT platform

Ledmotive manufacture spectrally-tunable digital lamps with different numbers of LED channels, sensor types and connectivity options. Each spectrally-tunable LED module provides feedback control through spectral and temperature sensors to account for wavelength shifts, flux fluctuations and includes overheating protection. The Ledmotive system includes a gateway with wireless connectivity sub-GHz, Wi-Fi, Bluetooth or DALI connectivity. The connected lighting platform allows customers to develop advanced lighting products that are intuitive and easy to use from a smartphone or directly through the cloud.

### 4. Standardized communication interfaces for smart lighting

Connectivity through a standardized wired or wireless communication interface is a key feature of a professional smart lighting system. However, for critical infrastructures and street lighting systems it is mandatory to use wired interfaces mainly based on DALI, power line communications and Ethernet due to the robustness of the communication interface.

Home lighting applications and illumination for non-critical infrastructures can use several standardized wireless communication interfaces. Smart lighting platforms prefer the use of sub-GHz, IEEE802.15.4, Bluetooth low energy, Wi-Fi, VLC or ZigBee Light Link connectivity in indoor environments. ZigBee Light Link and 6loWPAN use the same MAC and physical layer of the IEEE802.15.4 standard.

There are organizations promoting different wireless communication protocols (e.g., M2M Global Alliance, Internet of Things Alliance, and IPSO Alliance). Table 1 shows several wireless communications technologies for intelligent smart lighting solutions which would be combined with some radio physical layers in different radio bands such as IEEE 802.11/b/g/n/a/ac (Wi-Fi), IEEE 802.15.4 [16] (ZigBee Light Link [17], 6LoWPAN, THREAD), IEEE802.15.1 (Bluetooth) or LoRaWAN.

Table 1. IoT Wireless communication technologies for smart lighting products

| Technology | Governing body | Network type | Frequency | Maximum Throughput |
|---|---|---|---|---|
| Wi-Fi | IEEE802.11a/b/g/n/ac/ad | WLAN | 2.4, 3.6, 5.6 GHz | 6-780 Mb/s 6.75 Gb/s @20GHz |
| Z-Wave | Z-Wave | Mesh | 868.42 MHz | 100 Kb/s |
| Bluetooth Class1 | Bluetooth (IEEE802.15.1) | WPAN | 2.4 -2.4835 GHz | 1 Mb/s |
| Bluetooth Smart (BLE) | IoT interconnect | WPAN | 2.4 -2.4835 GHz | 1 Mb/s |
| ZigBee | IEEE802.15.4 | Mesh | 2.4 -2.4835 GHz | 250kb/s |
| THREAD | IEEE802.15.4+6loWPAN | Mesh | 2.4 -2.4835 GHz | 251kb/s |
| RFID | ISO, IEC, ASTM, DASH7, EPCGlobal | P2P | 13.56 MHz etc. | 423kb/s |
| **NFC** | ISO/IEC1357 etc. | P2P | 13.56 MHz | 424kb/s |
| **GPRS (2G)** | 3GPP | GERAN: GSM EDGE Radio Access Network | GSM 900 MHz and 1800 MHz | 171 kb/s |
| **EDGE (2G)** | 3GPP | GERAN | GSM 900 MHz and 1800 MHz | 384 kb/s |
| **UMTS (3G) HSDPA/HSUPA** | 3GPP | UTRAN: Universal Terrestrial Radio Access Network | 900-2100 MHz Europe | 0.73 -56 Mb/s |
| **LTE (4G)** | 3GPP | GERAN/UTRAN | 1800-2600 MHz Europe | 0.1-1 Gb/s |
| **ANT+** | ANT+ alliance | WSN | 2.4 GHz | 1 Mb/s |
| **Cognitive Radio** | IEEE802.22 WG | WRAN | 470-790 MHz | 24 Mb/s |
| **Weightless-N/W** | Weightless-SIG | LPWAN | ISM 868 MHz | 0.001-10 Mb/s |
| **6loWPAN** | IEEE802.15.4 | Mesh | 2.4 GHz | 255 kb/s |
| **LoRaWAN** | LoRa Alliance | LPWAN | 433/868/780/915 MHz | EU: 0.3 kbp/s to 50kbp/s  US: 0.9 kbp/s to -100 kbp/s |

Interoperability [18] in lighting appliances is being addressed toward standardized Internet protocols (IPv4 and IPv6). OpenAIS, a European project, is working in a new IoT architecture for lighting applications to increase the interoperability of heterogeneous lighting systems [19]. This solution is IP-based, extensible, interoperable, and secure; also it has been set up with key players from the lighting industry and IoT. This system defines an open architecture for lighting systems with standardized open APIs which also make the system interoperable with building automation systems and agnostic cloud services.

Additional agnostic schemes to increase the interoperability are based on IP standardization [20], that is a practical solution to alleviate communication problems between heterogeneous lighting devices with manufacturer dependent protocols or in some cases, closed ecosystems. In this case, the interoperability could be achieved by combining IPv4 and IPv6 standardization schemes with the inclusion of lightweight application protocols such as COAP and MQTT [21] running over UDP and TCP transport protocols respectively.

Finally, when energy constraints are imposed for efficient lighting devices, the implementation of lightweight protocols such as Bluetooth low energy (BLE) 5.0 or ZigBee light link would be a suitable solution. While for outdoor LED luminaires that are managed remotely, connectivity options are mainly covered with traditional wired protocols such as power line communications (PLC) or DALI.

## 4.1 Visible light communications

The modern concept of VLCs, also known as Li-Fi [22], concerns the technology aiming to transmit data while providing artificial light for illumination. This concept usually employs LEDs to illuminate and provide data for unidirectional or bidirectional communications. It means that information is encoded in the visible light spectrum (380 nm − 780 nm) transmitted by those LEDs, and the simplest way to do it is through digital modulation techniques [23].

As LED emitters are inexpensive, easy to fabricate and may be modulated at very high speeds in the range of several MHz; these advantages gives a chance for further applications mainly in optical communications such as VLC [24] and Li-Fi, and not only for general lighting or signaling applications. Digital lighting appliances prepared with optical links in the visible range have different advantages mainly in regards to the low interferences and high security they present, thanks to the fact that visible light does not pass through walls. Moreover, synchronization between luminaries with VLC interfaces may be achieved in order to allow a stable data transmission even if the receiver is moving in an indoor space. Nowadays, white light LED luminaires are the core of modern VLC solutions. That is why both research institutes and the industry are working on fast and energy efficient LEDs providing white light. There are different ways to provide white light using LEDs: with RGB LEDs, phosphorous coating (PC) LEDs and finally with OLEDs. Table 2 summarizes the main features and fundamentals of current state-of-the-art LED and OLED emitters for VLC applications.

Table 2. LEDs providing white light for VLC connectivity

| LED | Material type | Characteristics | Bandwidth | Application |
|---|---|---|---|---|
| **RGB LED** | Inorganic | Three different inorganic LEDs (red, green and blue) | 10-20 MHz | General illumination, LI-FI, LED-ID, optical camera communications |
| **PC-LED** | Inorganic | Blue LED made of InGaN with a phosphorous coating made of YAG that is a synthetic crystalline material | 3-5 MHz | General illumination, LI-FI, LED-ID, optical camera communications |
| **OLED** | Organic | Contains different layers and between the anode and cathode are two layers made from organic molecules called the emissive layer | <= 1 MHz | Indoor lighting, display |
| **µLED** | | Based on inorganic material such as GaN. MicroLEDs are extremely small, typically 1/10th the width of a human hair | ➢ 300 MHz | Wearable technology, mobile devices |

Experimental VLC systems are evolving very fast with optical systems with high throughputs [25] for location services. Test beds are intended for short range links, less than 2m, between the source and receiver. These schemes are being designed for indoor location purposes and multimedia services, supporting an average constant light intensity to avoid annoying flickering.

## 5. Conclusion

Smart lighting is expected to have huge impact during the upcoming years, due to the accelerated deployment of LED drivers, sensors and connected LED platforms. Behind this tendency, different vendors are in a race to connect smart LED luminaires on the same infrastructure of the IoT. This trend is initially being spread across home and professional applications in indoor and outdoor environments.

Lighting appliances and the IoT ecosystem converge in several areas: health and wellness with circadian LED systems, lighting systems with advanced sensing, optical communications, and location services.

Moreover, such lighting systems require the use of wired and wireless connectivity to be connected to the Internet. However, as the jungle of wireless standards and protocols is evolving, manufacturers need to adapt their products to lighting market tendencies especially when it is combined with the IoT ecosystem and advanced lighting control systems.

Finally, interoperable solutions for smart lighting are being explored mainly at the application level using REST web services. As these services offer a flexible solution to share data and commands with mobile apps and the cloud in different lighting ecosystems oriented to the Internet.